\documentclass{llncs}
\usepackage[T1]{fontenc}
\usepackage[latin9]{inputenc}
\usepackage{amssymb}
\usepackage{amsmath}
\usepackage{graphicx}

\usepackage[english]{babel}
\makeatletter
\newtheorem{thm}{\protect\theoremname}
  \newtheorem{prop}[thm]{\protect\propositionname}
  \newtheorem{defn}[thm]{\protect\definitionname}

\newcommand{\coeff}{\mathsf{coeff}}
\renewcommand{\mp}{\mathsf{F}}
\newcommand{\MSOL}{\mathrm{MSOL}}
\newcommand{\FPT}{\mathrm{FPT}}
\newcommand{\FPPT}{\mathrm{FPPT}}
\newcommand{\NP}{\mathrm{NP}}
\newcommand{\Wone}{\mathrm{W}}
\newcommand{\sharpP}{\mathrm{\sharp P}}
\newcommand{\qr}{\mathrm{qr}}
\newcommand{\N}{\mathbb{N}}
\newcommand{\Op}{\mathsf{Op}}

\makeatother

\usepackage{babel}
  \providecommand{\definitionname}{Definition}
  
  \providecommand{\propositionname}{Proposition}
  \providecommand{\remarkname}{Remark}
\providecommand{\theoremname}{Theorem}

\begin{document}

\title{Efficient computation of generalized Ising polynomials on graphs
with fixed clique-width\thanks{Tomer Kotek was supported by the Austrian
National Research Network S11403-N23 (RiSE) of the Austrian Science
Fund (FWF) and by the Vienna Science and Technology Fund (WWTF)
grant PROSEED. }}
\author{Tomer Kotek\inst{1} \and Johann A. Makowsky\inst{2}}
\institute{TU Vienna \\ Vienna, Austria \\ 
\email{tkotek@tuwien.ac.at} \and
Technion --- Israel Institute of Technology \\ Haifa, Israel \\ \email{janos@cs.technion.ac.il}}

\maketitle
\begin{abstract}
Graph polynomials which are definable in 
Monadic Second
Order Logic ($\MSOL$) on the vocabulary
of graphs are Fixed-Parameter Tractable ($\FPT$) with respect to clique-width.
In contrast, graph polynomials which are definable in $\MSOL$ on the vocabulary of hypergraphs are 
fixed-parameter tractable
with respect to tree-width, but not necessarily
with respect to clique-width. No algorithmic meta-theorem is known
for the computation of graph polynomials definable in $\MSOL$ on
the vocabulary of hypergraphs with respect to clique-width. We define an infinite class of such graph
polynomials extending the class of graph polynomials definable in
$\MSOL$ on the vocabulary of graphs and prove that they are Fixed-Parameter 
Polynomial Time ($\FPPT$) computable, i.e. that they can be computed
in time $O(n^{f(k)})$, where $n$ is the number of vertices and $k$
is the clique-width. 
\end{abstract}

\section{Introduction}

In recent years there has been growing interst in graph polynomials,
functions from graphs to polynomial rings which are invariant under
isomorphism. Graph polynomials encode information about the graphs
in a compact way in their evaulations, coeffcients, degree and roots.
Therefore, efficient computation of graph polynomials has received
considerable attention in the literature. Since most graph polynomials
which naturally arise are $\sharpP$-hard to compute (see e.g. \cite{valiant1979complexity,ar:JVW90,pr:BH08}),
a natural perspective under which to study the complexity of graph polynomials
is that of {\em parameterized complexity}. 

Parameterized complexity is a successful approach to tackling $\NP$-hard
problems \cite{bk:DF99,bk:FG06}, by measuring complexity with respect
to an additional {\em parameter} of the input; we will be interested in the parameters
tree-width and clique-width. A computational problem is {\em\bf fixed-parameter tractable}
($\FPT$) with respect to a parameter $k$ if it can be solved in
time $f(k)\cdot p(n)$, where $f$ is a computable function of $k$,
$n$ is the size of the input, and $p(n)$ is a polynomial in $n$.
Many $\NP$-hard problems are fixed parameter tractable for an appropriate
choice of parameter, see \cite{bk:FG06} for many examples. Every
problem in the infinite class of decision problems definable in {\em Monadic Second Order Logic}
($\MSOL$) is fixed-parameter tractable with respect to tree-width
by Courcelle's Theorem \cite{ar:Courcelle90,ar:ArnborgEtAl,bk:CourcelleEngelfriet12}
(though the result originally was not phrased in terms of parameterized
complexity). 

The computation problem we consider for a graph polynomial $P(G;x_{1},\ldots,x_{r})$
is the following:

~~~~~~~~~~%
\begin{minipage}[t]{1\columnwidth}%
$P-\mathrm{Coefficients}$

$\mathit{Instance}$: A graph $G$

$\mathit{Problem}:$ Compute the coefficients $a_{i_{1},\ldots,i_{r}}$
of the monomials $x_{1}^{i_{1}}\cdots x_{r}^{i_{r}}$. %
\end{minipage}

For graph polynomials, a parameterized complexity theory with respect
to tree-width has been developed. Here, the goal is to compute, given
an input graph, the table of coefficients of the graph polynomial.
The Tutte polynomial has been shown to be fixed-parameter tractable
\cite{ar:noble98,ar:andrzejak98}. \cite{makowsky2005coloured} used
a logical method to study the parameterized complexity of an infinite
class of graph polynomials, including the Tutte polynomial, the matching
polynomial, the independence polynomial and the Ising polynomial.
\cite{makowsky2005coloured} showed that the class of graph polynomials
definable in $\MSOL$ in the vocabulary of hypergraphs\footnote{In \cite{bk:CourcelleEngelfriet12}, $\MSOL$ in the vocabulary of hypergraphs is denoted $MS_2$, while $\MSOL$ in the vocabulary of graphs is denoted $MS_1$.}
is fixed-parameter tractable. This class contains the vast majority
of graph polynomials which are of interest in the literature. 

Going beyond tree-width to clique-width the situation becomes more
complicated. \cite{ar:CMR2000} studied the class of graph polynomials
definable in $\MSOL$ {\em in the vocabulary of graphs}. They proved
that every graph polynomial in this class is fixed-parameter tractable
with respect to clique-width. However, this class of graph polynomials
does not contain important examples such as the chromatic polynomial,
the Tutte polynomial and the matching polynomial. In fact, \cite{ar:FominGolovachLokshtanov10}
proved that the chromatic polynomial and the Tutte polynomial are
not fixed-parameter tractable with respect to clique-width (under
the widely believed complexity-theoretic assumption that $\FPT\not=\Wone[1]$).
\cite{makowsky2006computing} proved that the chromatic polynomial
and the matching polynomial are {\em\bf fixed-parameter polynomial time}
computable with respect to clique-width, meaning that they can be
computed in time $n^{f(cw(G))}$, where $n$ is the size of the graph,
$cw(G)$ is the clique-width of the graph and $f$ is a computable
function. \cite{kotek2012complexity} proved an analogous result for
the Ising polynomial. The main result of this paper is a meta-theorem
generalizing the fixed-parameter polynomial time computability of
the chromatic polynomial, the matching polynomial and the Ising polynomial
to an infinite family of graph polynomials definable in $\MSOL$ analogous
to \cite{ar:CMR2000}. 
\begin{thm}
Let $P$ be an $\MSOL$-Ising polynomial. $P$ is fixed-parameter polynomial
time computable with respect to clique-width. 
\end{thm}
The class of $\MSOL$-Ising polynomials is defined in Section \ref{se:MSOLISing}.

\section{Preliminaries}

Let $[k]=\left\{ 1,\ldots,k\right\}$. Let $\tau_{G}$ be the vocabulary
of graphs $\tau_{G}=\left\langle \mathbf{E}\right\rangle $ consisting
of a single binary relation symbol $\mathbf{E}$. A $k$-graph is a structure $(V,E,R_1,\ldots,R_k)$ which consists of a simple graph $G=(V,E)$ together with a partition $R_1,\ldots,R_k$ of $V$.  Let
$\tau_{G}^{k}$ denote the vocabulary of $k$-graphs $\tau_{G}^{k}=\left\langle \mathbf{E},\mathbf{R}_{1},\ldots,\mathbf{R}_{k}\right\rangle $
extending $\tau_{G}$ with unary relation symbols $\mathbf{R}_{1},\ldots,\mathbf{R}_{k}$. 

The class $CW(k)$ of $k$-graphs of clique-width at most $k$ is defined inductively. Singletons belong to $CW(k)$, and $CW(k)$ is closed under disjoint union $\sqcup$ and 
two other operations, $\rho_{i\to j}$ and $\eta_{i,j}$, to be defined next. 
For any $i,j\in[k]$, $\rho_{i\to j}(G,\bar{R})$ is obtained by relabeling any vertex with label $R_i$ to label $R_j$. 
For any $i,j\in[k]$, $\eta_{i,j}(G,\bar{R})$ is obtained by adding all possible edges $(u,v)$ between members of $R_i$ 
and members of $R_j$. The clique-width of a graph $G$ is the minimal $k$ such that there exists a labeling $\bar{R}$ for 
which $(G,\bar{R})$ belongs to $CW(k)$. We denote the clique-width of $G$ by $cw(G)$.
The clique-width operations $\rho_{i\to j}$ and $\eta_{i,j}$ are well-defined for $k$-graphs.
The definitions of these operations extend naturally to structures $(V,E,S)$ which expand $k$-graphs with $S\subseteq v$. 

A $k$-expression is a term $\mathsf{t}$ which consists of singletons, disjoint unions $\sqcup$, relabeling $\rho_{i\to j}$ and edge creations $\eta_{i,j}$, which witnesses that the graph $val(\mathsf{t})$ obtained by performing the operations on the singletons is of clique-width at most $k$. Every graph of tree-width at most $k$ is of clique-width at most $2^{k+1}+1$, cf. \cite{ar:CourcelleOlariu2000}. While computing the clique-width of a graph is $\NP$-hard, S. Oum and P. Seymour showed that given a graph of clique-width $k$, finding a $(2^{3k+2}-1)$-expression is fixed parameter tractable with clique-width as parameter, cf. \cite{ar:Oum2005,ar:SeymourOum2006}. 

For a formula $\varphi$, let $\qr(\varphi)$ denote the quantifier
rank of $\varphi$. For every $q\in\mathbb{N}$ and vocabulary $\tau$,
we denote by $\MSOL^{q}(\tau)$ the set of $\MSOL$-formulas on the
vocabulary $\tau$ which have quantifier rank at most $q$. For two
$\tau$-structures $\mathcal{A}$ and $\mathcal{B}$, we write $\mathcal{A}\equiv^{q}\mathcal{B}$
to denote that $\mathcal{A}$ and $\mathcal{B}$ agree on all the
sentences of quantifier rank $q$. 

\begin{definition}[Smooth operation]
 An $\ell$-ary operation $\Op$ on $\tau$-structures is called {\em smooth} 
if for all $q \in \N$, 
whenever  $\mathcal{A}_j \equiv^q \mathcal{B}_j$ for all $1\leq j\leq \ell$, 
we have  $$ \Op(\mathcal{A}_1,\ldots,\mathcal{A}_\ell) \equiv^q  
\Op(\mathcal{B}_1,\ldots,\mathcal{B}_\ell)\,.$$
\end{definition}

Smoothness
of the clique-width operations is an important technical tool for
us:
\begin{thm}
[Smoothness, cf. \cite{makowsky2004algorithmic}] \label{th:smooth}~
\begin{enumerate}
\item For every vocabulary $\tau$, the disjoint union $\sqcup$ of two
$\tau$-structures is smooth. 
\item For every $1\leq i\not=j\leq k$, $\rho_{i\to j}$ and $\eta_{i,j}$
are smooth. 
\end{enumerate}
\end{thm}
It is convenient to reformulate Theorem \ref{th:smooth} in terms
of {\em Hintikka sentences} (see \cite{bk:EF2005}):
\begin{prop}
[Hintikka sentences] \label{prop:Hin}  Let $\tau$ be a vocabulary.
For every $q\in\N$ there is a finite set
\[
\mathcal{H}_{\tau}^{q}=\{h_{1},\ldots,h_{\alpha}\}
\]
 of $\MSOL^{q}(\tau)$-sentences such that
\begin{enumerate}
\item Every $h\in\mathcal{H}_{\tau}^{q}$ has a finite model.
\item The conjunction $h_{1}\land h_{2}$ of any two distinct $h_{1},h_{2}\in\mathcal{H}_{\tau}^{q}$
is unsatisfiable.
\item Every $\MSOL^{q}(\tau)$-sentence $\theta$ is equivalent to exactly
one finite disjunction of sentences in $\mathcal{H}_{\tau}^{q}$.
\item Every $\tau$-structure $\mathcal{A}$ satisfies a unique member $hin_\tau^{q}(\mathcal{A})$
of $\mathcal{H}_{\tau}^{q}$.
\end{enumerate}
\end{prop}
In order to simplify notation we omit the subscript $\tau$ in $hin_\tau^{q}$ when $\tau$ is clear from the context.

Let $\tau_{{\bf S}}$ the be the vocabulary consisting of the binary relation
symbol ${\bf E}$ and the unary relation symbol ${\bf S}$.
Let $\tau_{{\bf S},k}$ extend $\tau_{{\bf S}}$ with the unary relation
symbols $\mathbf{R}_{1},\ldots,{\bf R}_{k}$. 
From Theorem \ref{th:smooth} and Proposition \ref{prop:Hin} we get:
\begin{thm}
For every $k\in \mathbb{N}^+$:
\begin{enumerate}
\item There is $\mp_{\sqcup}:\mathcal{H}_{\tau_{{\bf S},k}}^{q}\times\mathcal{H}_{\tau_{{\bf S},k}}^{q}\to\mathcal{H}_{\tau_{{\bf S},k}}^{q}$
such that, for every $\mathcal{M}_{1}$ and $\mathcal{M}_{2}$, 
\newline
$\mp_{\sqcup}(hin^{q}(\mathcal{M}_{1}),hin^{q}(\mathcal{M}_{2}))=hin^{q}(\mathcal{M}_{1}\sqcup\mathcal{M}_{2})$. 
\item For every unary operation $\mathsf{op}\in\{\rho_{p\to q},\eta_{p,q}:p,q\in[k]\}$,
there is $\mp_{\mathsf{op}}:\mathcal{H}_{\tau_{{\bf S},k}}^{q}\to\mathcal{H}_{\tau_{{\bf S},k}}^{q}$
such that, for every $\mathcal{M}$, $\mp_{\mathsf{op}}(hin^{q}(\mathcal{M}))=hin^{q}(\mathsf{op}(\mathcal{M}_{1}\sqcup\mathcal{M}_{2}))$. 
\end{enumerate}
\end{thm}

\subsection{$\MSOL$-Ising polynomials \label{se:MSOLISing}}

For every $t\in\mathbb{N}^{+}$, let $\tau_{t}=\tau\cup\{\mathbf{S}_{1},\ldots,{\bf S}_{t}\}$,
where ${\bf S}_{1},\ldots,{\bf S}_{t}$ are new unary relation symbols.
\begin{defn}
[$\MSOL$-Ising polynomials] \label{def:MSOLIsing} For every $t\in\N^{+}$,
$\theta\in\MSOL(\tau_{t})$ and $G=(V,E)$ we define $P_{t,\theta}(G;\bar{X},\bar{Y})$
as follows: 
\[
P_{t,\theta}(G;\bar{X},\bar{Y})=\sum_{\substack{S_{1}\sqcup\cdots\sqcup S_{t}=V:\\
G\models\theta(S_{1},\ldots,S_{t})
}
}\,\prod_{i=1}^{t}X_{i}^{|S_{i}|}\prod_{1\leq i_{1}\leq i_{2}\leq t}Y_{i_{1},i_{2}}^{|(S_{i_{1}}\times S_{i_{2}})\cap E|}
\]
$P_{t,\theta}$ is the sum over partitions $S_{1},\ldots,S_{t}$ of
$V$ such that $\left(G,S_{1},\ldots,S_{t}\right)$ satisfies $\theta$
of the monomials obtained as the product of $X_{i}^{|S_{i}|}$ for
all $1\leq i\leq t$ and $Y_{i_{1},i_{2}}^{|(S_{i_{1}}\times S_{i_{2}})\cap E|}$
for all $1\leq i_{1}<i_{2}\leq t$. \end{defn}

\begin{example}
[Ising polynomial] The trivariate Ising polynomial $Z(G;x,y,z)$
is a partition function of the Ising model from statistical mechanics
used to study phase transitions in physical systems in the case of
constant energies and external field. $Z(G;x,y,z)$ is given by 
\[
Z(G;x,y,z)=\sum_{S\subseteq V}x^{|S|}y^{|\partial S|}z^{|E(S)|}
\]
where $\partial S$ denotes the set of edges between $S$ and $V\backslash S$,
and $E(S)$ denotes the set of edges inside $S$. $Z(G;x,y,z)$ was
the focus of study in terms of hardness of approximation in \cite{ar:GJP03}
and in terms of hardness of computation under the exponential time
hypothesis was studied in \cite{kotek2012complexity}. \cite{kotek2012complexity}
also showed that $Z(G;x,y,z)$ is fixed-parameter polynomial time
computable.

$\ensuremath{Z(G;x,y,z)}$ generalizes a bivariate Ising polynomial,
which was studied for its combinatorial properties in \cite{ar:AndrenMarkstrom2009}.
\cite{ar:AndrenMarkstrom2009} showed that $Z(G;x,y,z)$ contains
the matching polynomial, the van der Waerden polynomial, the cut polynomial,
and, on regular graphs, the independence polynomial and clique polynomial. 

The evaluation of $P_{2,\mathsf{true}}(G;X_{1},X_{2},Y_{1,1},Y_{1,2},Y_{2,2})$ 
at 
$X_1=x$, $X_2=1$, $Y_{1,1}=z$, $Y_{1,2}=y$ and $Y_{2,2}=1$
gives $Z(G;x,y,z)$ and therefore $Z(G;x,y,z)$ is an $\MSOL$-Ising polynomial.
\end{example}

\begin{example}
[Independence-Ising polynomial] The independence-Ising polynomial
$I_{Is}(G;x,y)$ is given by
\[
I_{Is}(G;x,y)=\sum_{\substack{S\subseteq V\\
S\mbox{ is an independent set}
}
}x^{|S|}y^{|\partial S|}
\]
$I_{Is}(G;x,y)$ contains the independence polynomial as the evaluation
$I(G;x)=I_{Is}(G;x,1)$. See the survey \cite{levit2005independence}
for a bibliography on the independence polynomial. The evaluation
$y=0$ is $I_{Is}(G;x,0)=(1+x)^{\mathit{iso}(G)}$, where $\mathit{iso}(G)$
is the number of isolated vertices in $G$. $I_{Is}(G;x,y)$ is an
evaluation of an $\MSOL$-Ising polynomial: 
\[
I_{Is}(G;x,y)=P_{2,\theta_{I}}(G;1,x,1,y,1)
\]
where $\theta_{I}(S)=\forall x\forall y\,\left({\bf E}(x,y)\to\left(\neg S_2(x)\lor\neg S_2(y)\right)\right)$. 
\end{example}

\begin{example}
[Dominating-Ising polynomial] The Dominating-Ising polynomial is
given by $D_{Is}(G;x,y,z)$ 
\[
D_{Is}(G;x,y,z)=\sum_{\substack{S\subseteq V\\
S\mbox{ is a dominating set}
}
}x^{|S|}y^{|\partial S|}z^{|E(S)|}
\]
where $\partial S$ denotes the set of edges between $S$ and $V\backslash S$.
$D_{Is}(G;x,y,z)$ contains the domination polynomial $D(G;x)$. $D(G;x)$
is the generating function of its dominating sets and we have $D_{Is}(G;x,1,1)=D(G;x)$.
The domination polynomial first studied in \cite{arocha2000mean}
and it and its variations have received considerable attention in
the literature in the last few years, see e.g. \cite{alikhani2009dominating,akbari2010zeros,akbari2010characterization,alaeiyan2011cyclically,DBLP:journals/combinatorics/KotekPSTT12,DBLP:journals/gc/KotekPT14,DBLP:journals/arscom/AkbariO14,alikhani2014introduction,brown2014roots,kahat2014dominating,DBLP:journals/dmgt/DodKPT15}.
Previous research focused on combinatorial properties such as recurrence
relations and location of roots. Hardness of computation was addressed
in \cite{kotek2013domination}. $D_{Is}(G;x,y,z)$ encodes the degrees
of the vertices of $G$: the number of vertices with degree $j$ is
the coefficient of $xy^{j}$ in $D_{Is}(G;x,y,z)$. $D_{Is}(G;x,y,z)$
is an $\MSOL$-Ising polynomial given by $P_{2,\theta_{D}}(G;x,1,z,y,1)$,
where 
\[\theta_{D}=\forall x\left(S_1(x)\lor\exists y\,\left(S_1(y)\land{\bf E}(x,y)\right)\right)\,.\]
\end{example}

\subsection{$\MSOL$-Ising polynomials vs $\MSOL$-polynomials}

Two classes of graph polynomials which have received attention in the literature are:
\begin{enumerate}
 \item $\MSOL$-polynomials on the vocabulary of graphs, and
 \item $\MSOL$-polynomials on the vocabulary of hypergraphs. 
\end{enumerate}
See e.g. \cite{ar:KM14connection} for the exact definitions. The former class contains graph polynomials such as the independence polynomial
and the domination polynomial. The latter class contains graph polynomials such as the Tutte polynomial and the matching polynomial. 
Every graph polynomial which is $\MSOL$-definable on the vocabulary of graphs is also $\MSOL$-definable on the vocabulary of hypergraphs. 

\begin{figure}
  \caption{\label{fig} Containments of classes of graph polynomials definable in $\MSOL$. }
  \ \\
  \centering
    \includegraphics{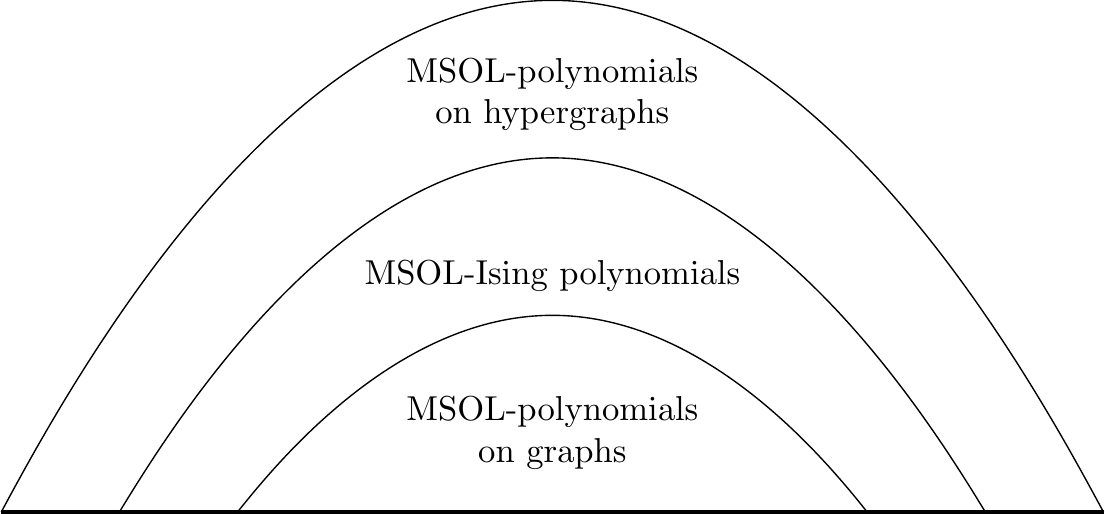}
\end{figure}

The class of $\MSOL$-Ising polynomials strictly contains the $\MSOL$-polynomials on graphs, see Fig. \ref{fig}. The containment is by definition. 
For the strictness, we use the fact that by definition the maximal degree of any indeterminate in an $\MSOL$-polynomial on graphs grows at most linearly in the number of vertices, while 
the maximal degree of $y$ in the Ising polynomial $Z(K_{n,n};x,y,z)$
of the complete bipartite graph $K_{n,n}$ equals $n^2$.

Every $\MSOL$-Ising polynomial $P_{t,\theta}$ is an $\MSOL$-polynomial on the
vocabulary of hypergraphs, given e.g. by 
\[
\sum_{\bar{S}}\sum_{\bar{B}}\,\prod_{i=1}^{t}X_{i}^{|S_{i}|}\prod_{1\leq i_{1}\leq i_{2}\leq t}Y_{i_{1},i_{2}}^{|B_{i_{1},i_{2}}|}
\]
where the summation over $\bar{S}$ is exactly as in Definition \ref{def:MSOLIsing},
and the summation over $\bar{B}$ is over tuples $\bar{B}=(B_{i_{1},i_{2}}:1\leq i_{1}\leq i_{2}\leq t)$
of subsets of the edge set of $G$ satisfying $\bigwedge_{i_{1},i_{2}}\psi_{i_{1},i_{2}}$,
where 
\[
\psi_{i_{1},i_{2}}=\forall x\forall y\left(B_{i_1,i_2}(x,y)\leftrightarrow \left({\bf E}(x,y)\land (S_{i_{1}}(x)\land S_{i_{2}}(y)\lor S_{i_{1}}(y)\land S_{i_{2}}(x))\right)\right)
\]
We use the fact that $S_1,\ldots,S_t$ is a partition of the set of vertices is definable in $\MSOL$.  

\section{Main result}

We are now ready to state the main theorem and prove a representative
case of it. 
\begin{thm}
[Main theorem]\label{th:main} For every $\MSOL$-Ising polynomial
$P_{t,\theta}$ there is a function $f(k,\theta,t)$ such that $P_{t,\theta}(G;\bar{X},\bar{Y},\bar{Z})$
is computable on graphs $G$ of size $n$ and of clique-width at most
$k$ in running time $O(n^{f(k,\theta,t)})$. 
\end{thm}

We prove
the theorem for graph polynomials of the form 
\[
Q_{\theta}(G;X,Y)=\sum_{S:\, G\models\theta(S)}X^{|S|}Y^{|\partial S|}
\]
 for every $\theta\in\MSOL(\tau_{{\bf S}})$. The summation in $Q_{\theta}$
is over subsets $S$ of the vertex set of $G$. The graph polynomials
$Q_{\theta}$ are a notational variation of $P_{t,\theta}$ with $t=2$, $X_2=1$ and $Y_{1,1}=Y_{2,2}=1$: 
for every $\theta\in\MSOL(\tau_{2})$, $P_{2,\theta}(G;X,1,1,Y,1)=Q_{\theta'}(G;X,Y)$,
where $\theta'$ is obtained from $\theta$ by substituting $\mathbf{S}_{1}$
with $\mathbf{S}$ and $\mathbf{S}_{2}$ with $\neg\mathbf{S}$. The proof for the general case is 
in similar spirit. 

For every $q\in\mathbb{N}$ there is a finite set $\mathfrak{A}_{q}$ of $\MSOL(\tau_{\mathbf{S},k})$-Ising polynomials
 such that, for every formula $\theta\in \MSOL^q(\tau_{\mathbf{S}})$,
$Q_{\theta}$ is a sum of members of $\mathfrak{A}_{q}$ (see below).  
The algorithm computes the values of the members of $\mathfrak{A}_{q}$ on  $G$
by dynamic programming over the parse term of $G$, and using those values, the value of $Q_\theta$ on $G$.   

More precisely, for every $\beta\in\mathcal{H}_{\tau_{{\bf S},k}}^{q}$,
let 
\[
A_{\beta}(G;\bar{x},\bar{y})=\sum_{S:\, G\models\beta(S)}\,\prod_{1\leq c\le k}x_{c}^{|S\cap R_{c}|}\prod_{1\leq c_{1},c_{2}\leq k}y_{c_{1},c_{2}}^{|(R_{c_{1}}\cap S)\times(R_{c_{2}}\backslash S)|}
\]
and let 
$$\mathfrak{A}_q=\{A_{\beta}:\beta\in\mathcal{H}_{\tau_{{\bf S},k}}^{q}\}\,.$$
Every $\theta\in\MSOL^{q}(\tau_{\mathbf{S}})$ also belongs to $\MSOL^{q}(\tau_{\mathbf{S},k})$, and hence there
exists by Proposition \ref{prop:Hin} a set $\mathcal{H}\subseteq\mathcal{H}_{\tau_{{\bf S},k}}^{q}$
such that 
\[
\theta\equiv\bigvee_{h\in\mathcal{H}}h
\]
Hence,
\begin{equation}
Q_{\theta}(G;X,Y)=\sum_{h\in\mathcal{H}}A_{h}(G;\bar{x},\bar{y})\label{eq:sumQA}
\end{equation}
 setting $x_{c}=X$ and $y_{c_{1},c_{2}}=Y$ for all $1\leq c,c_{1},c_{2}\leq k$.

For tuples $\bar{b}=\left((b_{c}:c\in[k]),(b_{c_{1},c_{2}}:c_{1},c_{2}\in[k])\right)\in[n]^{k}\times[n]^{k^2}$,
let $\coeff_{\theta}^{G}(\bar{b})\in\mathbb{N}$ be the coefficient
of 
\[
\prod_{c}x_{c}^{b_{c}}\prod_{c_{1},c_{2}}y_{c_{1},c_{2}}^{b_{c_{1},c_{2}}}
\]
in $A_{\beta}(G;\bar{x},\bar{y})$.

\subsection*{Algorithm.}

Given a $k$-graph $G$, the algorithm first computes a parse tree
$\mathsf{t}$ as in \cite{ar:Oum2005,ar:SeymourOum2006}. The algorithm
then computes $A_{\beta}(G;\bar{x},\bar{y})$ for all $\beta\in\mathcal{H}_{\tau_{{\bf S},k}}^{q}$
by induction over $\mathsf{t}$:
\begin{enumerate}
\item If $G$ is a graph of size $1$, then $A_{\beta}(G)$ is computed
directly. 
\item Let $G$ be the disjoint union of $H_{A}$ and $H_{B}$. We compute
$\coeff_{\beta}^{G}(\bar{b})$ for every $\beta\in\mathcal{H}_{\tau_{\mathbf{S},k}}^{q}$
and $\bar{b}\in[n]^{k}\times[n]^{k^2}$ as follows: 
\[
\coeff_{\beta}^{G}(\bar{b})=
\sum_{h_{1},h_{2}:\mp_{\sqcup}(h_{1},h_{2})\models\beta}\sum_{\bar{d}+\bar{e}=\bar{b}}\coeff_{\beta}^{H_{A}}(\bar{d})\coeff_{\beta}^{H_{B}}(\bar{e})
\]
 
\item Let $G=\rho_{p\to q}(H)$. We compute $\coeff_{\beta}^{G}(\bar{b})$
for every $\beta\in\mathcal{H}_{\tau_{{\bf S},k}}^{q}$ and $\bar{b}\in[n]^{k}\times[n]^{k^2}$
as follows: 
\[
\coeff_{\beta}^{G}(\bar{b})=\sum_{h:\mp_{\rho_{p\to q}}(h)\models\beta}\sum_{\bar{d}}\coeff_{h}^{H}(\bar{d})
\]
where the inner summation is over $\bar{d}$ such that 
\[
b_{c}=\begin{cases}
d_{c} & c\notin\{p,q\}\\
d_{p}+d_{q} & c=q\\
0 & c=p
\end{cases}
\]
and 
\[
b_{c_{1},c_{2}}=\begin{cases}
d_{c_{1},c_{2}} & c_{1},c_{2}\notin\{p,q\}\\
0 & p\in\{c_{1},c_{2}\}\\
d_{q,q}+d_{p,p}+d_{p,q}+d_{q,p} & c_{1}=c_{2}=q\\
d_{q,c_{2}}+d_{p,c_{2}} & c_{1}=q,c_{2}\not\in\{q,p\}\\
d_{c_{1},q}+d_{c_{1},p} & c_{2}=q,c_{1}\not\in\{q,p\}
\end{cases}
\]

\item Let $G=\eta_{p,q}(H)$ with $p\not=q$. Let $n_G$ be the number of vertices in $G$. 
We compute $\coeff_{\beta}^{G}(\bar{b})$
for every $\beta\in\mathcal{H}_{\tau_{{\bf S},k}}^{q}$ and $\bar{b}\in[n]^{k}\times[n]^{k^2}$
as follows: 
\[
\coeff_{\beta}^{G}(\bar{b})=\sum_{h:\mp_{\eta_{p,q}}(h)\models\beta}\sum_{\bar{d}}\coeff_{h}^{H}(\bar{d})
\]
where the summation is over $\bar{d}$ such that $b_{c}=d_{c}$ and
\[
b_{c_{1},c_{2}}=\begin{cases}
d_{c_{1},c_{2}} & \{c_{1},c_{2}\}\not=\{p,q\}\\
d_{p}(n_G-d_{q}) & c_1 = p,\,c_2=q\\
d_{q}(n_G-d_{p}) & c_1 = q,\,c_2=p
\end{cases}
\]

\end{enumerate}

Finally, the algorithm computes $Q_{\theta}$
as the sum from Eq. (\ref{eq:sumQA}).
\subsection{Runtime}

The main observations for the runtime analysis are:
\begin{itemize}
\item The size of the set $\mathcal{H}_{\tau_{{\bf S},k}}^{q}$ of Hintikka
sentences is a function of $k$ but does not depend on $n$. Let $\mathsf{s}_{\tau_{{\bf S},k}}^{q}=|\mathcal{H}_{\tau_{{\bf S},k}}^{q}|$. 
\item By definition of $A_{\beta}$, for a monomial $\prod_{1\leq c\le k}x_{c}^{i_{c}}\prod_{1\leq c_{1},c_{2}\leq k}y_{c_{1},c_{2}}^{j_{c_{1},c_{2}}}$
to have a non-zero coefficient, it must hold that $i_{c}\leq n$ and
$j_{c_{1},c_{2}}\le\binom{n}{2}$, since $i_{c}$ and $j_{c_{1},c_{2}}$
are sizes of sets of vertices and sets of edges, respectively.
\item The coefficient of any monomial of $A_{\beta}$ is at most $2^{n}$. 
\item The parse tree guaranteed in \cite{ar:Oum2005,ar:SeymourOum2006}
is of size $O(n^{c}f_{1}(k))$ for suitable $f_{1}$ and $c$. 
\end{itemize}
The algorithm performs a single operation for every node of the parse
tree. 

\textbf{Singletons}: the coefficients of every $A_\beta\in\mathfrak{A}_q$ for a singleton $k$-graph
can be computed in time $O(k)$, which can be bounded by $O(n^{k})$. 

\textbf{Disjoint union},\textbf{ recoloring} \textbf{and} \textbf{edge
additions}: the algorithm sums over (1) $h\in\mathcal{H}_{\tau_{{\bf S},k}}^{q}$
or pairs $(h_{1},h_{2})\in\left(\mathcal{H}_{\tau_{{\bf S},k}}^{q}\right)^{2}$
and (2) over $\bar{d}\in[n]^{k}\times[n]^{k^2}$ or pairs
$(\bar{d},\bar{e})\in\left([n]^{k}\times[n]^{k^2}\right)^{2}$,
then (3) performs a fixed number of arithmetic operations on numbers
which can be written in $O(n)$ space. 

Each node in the parse tree requires time at most $O\left(n^{k}(\mathsf{s}_{\tau_{{\bf S},k}}^{q})^{2}\left([n]^{k}\times[n]^{k^2}\right)^{2}\right)$.
Since the size of the parse tree is $O(n^{c}f_{1}(k))$, the algorithm
runs in fixed-parameter polynomial time.

\section{Conclusion}

We have defined a new class of graph polynomials, the $\MSOL$-Ising
polynomials, extending the $\MSOL$-polynomials on the vocabulary
of graphs and have shown that every $\MSOL$-Ising polynomial can
be computed in fixed-parameter polynomial time. This result raises
the question of which graph polynomials are $\MSOL$-Ising polynomials.
In previous work \cite{pr:GKM08,pr:Makowsky09icla,ar:KM14connection}
we have developed a method based on connection matrices to show that graph
polynomials are not definable in $\MSOL$ over either the vocabulary
of graphs or hypergraphs. 
\begin{problem}
How can connection matrices be used to show that graph polynomials
are not $\MSOL$-Ising polynomials?
\end{problem}
The Tutte polynomial does not seem to be an $\MSOL$-Ising polynomial.
\cite{gimenez2006computing} proved that the Tutte polynomial can
be computed in subexponential time for graphs of bounded clique-width. 
More precisely, the time bound in \cite{gimenez2006computing}
is of the form
$\exp(n^{1-f(cw(G))})$, where $0<f(i)<1$ for all $i\in \mathbb{N}$. 
\begin{problem}
Is there a natural infinite class of graph polynomials definable in
$\MSOL$ which includes the Tutte polynomial such that membership
in this class implies {\em fixed parameter subexponential time} computability
with respect to clique-width
(i.e., that the graph polynomial is computable in $\exp(n^{1-g(cw(G))})$
time for some function $g$ satisfiying $0<g(i)<1$ for all $i\in \mathbb{N}$)?
\end{problem}

\subsection*{Acknowledgement}
We are grateful to Nadia Labai for her comments and suggestions. 
\bibliographystyle{plain}

\end{document}